\documentclass[12pt]{article}
\usepackage[margin=1in]{geometry}  
\usepackage{setspace}
\usepackage[caption=false]{subfig}
\usepackage{epsfig}
\usepackage{graphics,graphicx,color}
\usepackage{amssymb,amsfonts,amsthm,amsmath}
\usepackage{multirow, booktabs}
\usepackage{mathrsfs}
\usepackage{natbib}
\usepackage{url}

\newcommand{\bm}[1]{\mbox{\boldmath$ #1 $\unboldmath}}

\newcommand{\diag}{\mbox{diag}}
\newcommand{\cov}{\mbox{cov}}

\newcommand{\var}{\mbox{var}}

\title{Bayesian Auxiliary Variable Model for Birth Records Data with Qualitative and Quantitative Responses}

\author{Xiaoning Kang$^{1}$, Shyam Ranganathan$^{2}$\footnote{Xiaoning Kang and Shyam Ranganathan share the first authorship due to equal contribution.},\\
Lulu Kang$^{3}$\footnote{Lulu Kang is the corresponding author.}, Julia Gohlke$^4$, Xinwei Deng$^2$}
\date{}
\begin{document}

\setstretch{1.3}
\maketitle
 \begin{center}
 $^1$Institute of Supply Chain Analytics and International Business College, \\
 Dongbei University of Finance and Economics, Dalian, China. \\ 
 Email: \url{kangxiaoning@dufe.edu.cn}\\
 $^2$Department of Statistics, Virginia Tech, Blacksburg, VA, U.S.A.\\
 Email: \url{shyam81@vt.edu} and \url{xdeng@vt.edu}\\
 $^3$Department of Applied Mathematics, Illinois Institute of Technology, Chicago, IL, U.S.A.\\
 Email: \url{lkang2@iit.edu}\\
 $^4$Department of Population Health Sciences, Virginia Tech, Blacksburg, VA, U.S.A\\
 Email: \url{jgohlke@vt.edu}
 \end{center}

\begin{abstract}
Many applications involve data with qualitative and quantitative responses.
When there is an association between the two responses, a joint model will provide improved results than modeling them separately.
In this paper, we propose a Bayesian method to jointly model such data.
The joint model links the qualitative and quantitative responses and can assess their dependency strength via a latent variable.
The posterior distributions of parameters are obtained through an efficient MCMC sampling algorithm.
The simulation shows that the proposed method can improve the prediction capacity for both responses.
We apply the proposed joint model to the birth records data acquired by the Virginia Department of Health and study the mutual dependence between preterm birth of infants and their birth weights. 
\end{abstract}

\newpage

\section{Introduction}

In many applications, mutually dependent quantitative and qualitative (QQ) types of outcome data are simultaneously observed.
It is important to jointly model them to make accurate estimations and inferences, which provide scientific and meaningful conclusions.
In this paper, our application focuses on a birth records study examining the mutual dependency of birth weight and preterm birth.
The birth weight of the infant, a quantitative outcome, is an important variable that doctors need to monitor \citep{LBWtrends}.
The average birth weight of healthy infants is about 3.5 kilograms.
Children with low birth weight are more likely to have complications soon after birth and later in life, compared to children with normal birth weights \citep{LBW1, shah2011intention}.
The birth weight is known to be related to another key variable, \emph{preterm birth}, a qualitative outcome variable whose value is set to 1 if an infant is born before 36 gestational weeks, and is 0 otherwise.
Several factors are related to low birth weight and preterm birth.
Such factors include the socio-economic and health status of the mother, stresses caused by the environment, etc. \citep{PTBoverview, PTB1}.
Naturally, a preterm born infant is more likely to suffer from low birth weight, and the two are highly correlated.
Both preterm birth and low birth weight are rare outcomes in the population, accounting for less than $10\%$ of all live births.
Meanwhile, they have a significant effect on the health of the population, as well as the economy in general, due to the expenses spent on caring and monitoring infants \citep{PTBcost}.

Modernized maternal care is designed to provide personalized healthcare to mothers and children.
It is important to understand how various factors affect both preterm birth and low birth weight.
A statistical model that accurately predicts both quantitative and qualitative outcomes may offer useful information to health practitioners and expectant mothers.
Many other applications are in need of such a joint model for quantitative and qualitative responses.
For example, in \cite{deng2015QQ} and \cite{kang2018bayesian}, the total thickness variation (continuous) and the site total indicator reading (binary) are both measured to evaluate the quality of the wafer after the lapping stage in the wafer manufacturing process.
In \cite{moustaki2000generalized}, survey data with both quantitative scores and categorical answers are jointly analyzed.
More applications and methodologies on the mixed type of quantitative and qualitative response data are reviewed in Section \ref{sec: literature}.

In this article, we develop a Bayesian hierarchical model for the mixed quantitative and qualitative types of responses.
We use a latent variable to connect the two types of responses, which is similar to the joint model introduced in \cite{catalano1992bivariate}.
This joint model is suitable for the data on the birth records study described previously. It is much simpler than the joint models by \cite{dunson2000bayesian}, and yet still sufficiently effective.
In \cite{catalano1992bivariate}, the joint distribution is factorized into two regression components--the marginal distribution of the continuous outcome and the conditional distribution of the binary outcome conditioned on the continuous outcome.
The latter is obtained through the latent variable, which is correlated with the observed continuous outcome.
Based on the factorization, the estimation is done in two steps.
The first step is to estimate the marginal regression model of the continuous outcome.
The second step is to estimate the probit regression model of the binary outcome conditioned on the continuous outcome.
The generalized estimating equations approach is used to obtain the estimation.
Different from \cite{catalano1992bivariate}, we incorporate the Bayesian framework, assume the proper prior and hyper prior distributions, derive the posterior distributions, and then develop the MCMC sampling procedure to obtain the posterior distributions.
Compared to the frequentist approach in \cite{catalano1992bivariate}, there are some merits with the Bayesian approach.
First, the posterior distribution of the latent variable is available.
Second, the Bayesian inference is more accurate since it is not based on the asymptotic distribution as in maximum likelihood estimation.
Third, sparsity on both regression models of the two outcomes is applied, due to the informative prior distributions, we assume for the regression coefficients, which is equivalent to the ridge regression.

The remainder of the article is outlined as follows. Section \ref{sec: literature} provides a literature of recent work on modeling quantitative-qualitative responses.
In Section \ref{sec:model}, we introduce the joint quantitative-qualitative model via latent variable in the Bayesian framework.
The full-conditional distributions of the parameters and the leave-one-out conditional posterior distribution of the latent variable.
Section \ref{sec:MCMC} lays out the MCMC sampling procedures.
Numerical study and the case study in birth records are provided in Section \ref{sec:simulation} and \ref{sec:app} to illustrate the merits of the proposed model.
This article concludes in Section \ref{sec:end}.

\section{Literature review} \label{sec: literature}

Some works in the literature have tackled the issue of mixed continuous and discrete types of outcomes.
Some of them, such as \cite{wang2007run, liu2014integration, cheng2015time, zhou2006bayesian, shi2006stream}, modeled the two types of responses separately.
They overlooked the possible association that may exist between the two types of responses.
As a result, if there exists a dependency between the two types of responses, separate modeling could lead to less accurate prediction and misinterpretation compared to the joint models.
Most other works are on joint models for mixed types of outcomes.
Such works include \cite{olkin1961multivariate, cox1992response, catalano1992bivariate, fitzmaurice1995regression, moustaki2000generalized, dunson2000bayesian, gueorguieva2001correlated, dunson2003dynamic, deng2015QQ, kang2018bayesian}.
Some interesting practical application cases can be found in \cite{zhou2006bayesian, mcculloch2008joint, hwang2014semiparametric, yeung2015bayesian, sun2017functional}.

These works can be further categorized into different groups.
From the perspective of estimation methods, these methods can be divided into Bayesian methods, such as \cite{dunson2000bayesian,dunson2003dynamic,kang2018bayesian}, and non-Bayesian methods, such as \cite{catalano1992bivariate, moustaki2000generalized, gueorguieva2001correlated, deng2015QQ}.
Depending the form of the joint model, methods such as in \cite{fitzmaurice1995regression, deng2015QQ, kang2018bayesian} considered modeling the quantitative response conditioned on the qualitative response, leading to conditional linear regression models and marginal classification models, whereas other methods such as \cite{catalano1992bivariate, dunson2000bayesian, moustaki2000generalized, gueorguieva2001correlated, dunson2003dynamic} used latent variable to link the discrete and continuous outcomes.

We highlight some representatives of the latent variable models.
Motivated to analyze a toxicity experiment, \cite{catalano1992bivariate} used a latent variable to obtain a joint distribution of mixed responses.
The joint distribution is a product of a linear regression model for the quantitative variable and a probit model for the qualitative variable.
\cite{dunson2000bayesian} used the generalized linear models to describe the joint distribution of variables and proposed a Markov chain Monte Carlo (MCMC) sampling algorithm for estimating the posterior distributions of the parameters.
\cite{dunson2003dynamic} extended the previous work to multidimensional longitudinal data.
However, such early methods focus on the model estimation without investigating a sparse and interpretable model.
Different from \cite{catalano1992bivariate}, the latent variables in \cite{dunson2000bayesian} and \cite{dunson2003dynamic} appear in the generalized linear model as the linear coefficients.
But in \cite{catalano1992bivariate}, the latent variable is used to define the probit model for the binary outcome.

\cite{deng2015QQ} proposed the QQ model for joint fitting quantitative and qualitative responses by the maximum likelihood estimation and identified the significant variables by imposing non-negative garrote constraints on the likelihood function.
The likelihood of the joint QQ model is the product of the conditional distribution of the quantitative responses conditioned on the qualitative responses and the marginal distribution of the qualitative responses.
The authors also developed an iterative algorithm to solve the constrained optimization problem.
Consequently, the classic asymptotical distribution of the maximum likelihood estimation cannot be easily applied, hence, making it difficult to conduct statistical inference.
Using the same QQ model in \cite{deng2015QQ} as the sampling distribution of the data, \cite{kang2018bayesian} introduced a sparse hierarchical Bayesian framework, which can easily provide statistical inference on the estimated parameters and prediction of the QQ model.
However, since they constructed their model by fitting the quantitative response conditioned on the qualitative response, \cite{deng2015QQ} and \cite{kang2018bayesian} appeared to improve the prediction accuracy for the quantitative response, while the model of qualitative response would be similar as it was modeled independently of the quantitative response.

\section{Bayesian QQ model with a latent variable}\label{sec:model}

\subsection{Sampling Distribution}\label{subsec:sampling}
Denote the observed data as $(\bm x_{i}, y_{i}, z_{i}), i = 1, \ldots, n$,  where $ y_{i} \in \mathbb{R}$ and $z_{i} \in  \{ 0, 1 \}$ are the continuous and binary observations respectively.
Here the predictor vector $\bm x = (x_{1}, \ldots, x_{p})'$ contains $p$ predictors (intercept is included if needed).
To jointly model the mixed-type of responses $Y$ and $Z$ given $\bm x$, the key is to describe the association between the two.
We introduce a latent variable of $U$ to facilitate this task.
Assume the binary response follows the Bernoulli distribution
\begin{align}\label{eq: model-1}
Z = \left \{
\begin{array}{cl}
1, &\textrm{if}\quad U \ge 0  \\
0, &\textrm{else if} \quad U <0
\end{array}
\right.
\mbox{ with } U|\bm \beta_1, \bm x \sim N(\bm x'\bm \beta_{1}, 1),
\end{align}
where $U$ is a latent variable for the binary response $Z$.
This kind of latent variable approach is also used in cases other than the mixed types of outcomes.
For example, \cite{holmes2006bayesian} used an auxiliary variable in Bayesian binary and multinomial regression.
Regarding the quantitative response $Y$, its marginal distribution is assumed to be
\begin{align} \label{eq: model-2}
Y |\bm \beta_2,\sigma^2, \bm x \sim N( \bm x'\bm \beta_{2},  \sigma^2).
\end{align}
To link the continuous and binary responses, we introduce a joint distribution of $(U, Y)$, and assume a bivariate normal distribution with parameters $\bm \theta=(\bm \beta_1,\bm \beta_2,\sigma^2,\rho)$.
\begin{align*}
\left.\left[
\begin{array}{c}
U \\
Y
\end{array}
\right]\right\vert \bm \theta, \bm x
\sim
N(\bm \mu, \bm \Sigma) \mbox{ with }
\bm \mu = \left[
\begin{array}{c}
\bm x' \bm \beta_{1} \\
\bm x' \bm \beta_{2}
\end{array}
\right],
\bm \Sigma = \left [
\begin{array}{cc}
1 & \rho  \sigma \\
\rho  \sigma & \sigma^2
\end{array}
\right].
\end{align*}
If $\rho$ is positive, meaning that $Y$ and the probability of $Z=1$ is positively correlated, then the larger the value of $Y$ the more likely that $Z$ would be equal to 1.
So to conclude the association between $Y$ and $Z$, the key is to estimate $\rho$ and make inference on the estimation.

\subsection{Full-conditional distributions}\label{subsec:post}

In this part, we detail the derivation of the posterior distributions for the parameters $p(\bm \theta|\bm y, \bm z,\bm X)$, where $\bm y=(y_1,\ldots, y_n)'$, $\bm z=(z_1,\ldots, z_n)'$ and
$\bm X$ is the model matrix of the regression with each row as $\bm x_i'$.
Based on the model assumption in Section \ref{subsec:sampling}, the joint distribution of $(Y, Z, U)$ can be directly written as follows, given a single point of input $\bm x$.
\begin{align*}
p(z=1, y, u|\bm \theta,\bm x) & = \Pr(Z=1|U=u)p(u,y|\bm \theta,\bm x)=I(u\geq 0)p(u,y|\bm \theta,\bm x),\\
p(z=0, y, u|\bm \theta,\bm x) & = \Pr(Z=0|U=u)p(u,y|\bm \theta,\bm x)=I(u<0)p(u,y|\bm \theta,\bm x),\\
p(z, y, u|\bm \theta,\bm x) & =[zI(u\geq 0)+(1-z)I(u< 0)]p(u,y|\bm \theta,\bm x).
\end{align*}
The joint sampling distribution of the two response variables is
\begin{align*}
p(z,y|\bm \theta,\bm x)
&=(1-z)p(y|\bm \theta,\bm x)+(2z-1)\int I(u\geq 0)p(u,y|\bm \theta,\bm x)du.
\end{align*}
To obtain the exact form of $\int I(u\geq 0)p(u,y|\bm \theta,\bm x)du$, we rewrite $p(u,y|\bm \theta,\bm x)$ into $p(u|y, \bm \theta,\bm x)p(y|\bm \theta,\bm x)$.
Based on the bivariate normal distribution of $(U,Y)$, the distribution $U|Y=y$ is
\begin{align*}
U|y, \bm \theta,\bm x \sim N \left(\bm x'\bm \beta_{1} +  \frac{\rho}{\sigma}(y-\bm x' \bm \beta_{2}), (1 - \rho^2) \right).
\end{align*}
So we obtain the following
\begin{align*}
&\int I(u\geq 0)p(u,y|\bm \theta,\bm x)du=\int I(u\geq 0)p(u|y,\bm \theta,\bm x)p(y|\bm \theta,\bm x)du\\
=&p(y|\bm \theta,\bm x)\Phi\left(\left.\frac{\bm x'\bm \beta_1+\frac{\rho}{\sigma}(y-\bm x'\bm \beta_2)}{\sqrt{(1 - \rho^2) }}\right\vert y, \bm \theta,\bm x\right).
\end{align*}
Here $\Phi(\cdot)$ represents the CDF of the standard normal random variable.
To simplify the notation, define
\[
s(y|\bm \theta,\bm x)=\frac{\bm x'\bm \beta_1+\frac{\rho}{\sigma}(y-\bm x'\bm \beta_2)}{\sqrt{(1 - \rho^2) }}.
\]
Thus, the joint distribution of $(Z,Y)$ can be written as
\begin{equation*}
p(z,y|\bm \theta,\bm x)=p(y|\bm \theta,\bm x)\left[(1-\Phi(s(y)|\bm \theta,\bm x))+z(2\Phi(s(y)|\bm \theta,\bm x)-1)\right],
\end{equation*}
or more explicitily
\begin{align*}
& p(z=1,y|\bm \theta,\bm x)=p(y|\bm \theta,\bm x)\Phi(s(y)|\bm \theta,\bm x), \quad p(z=0, y|\bm \theta,\bm x)=p(y|\bm \theta,\bm x)(1-\Phi(s(y)|\bm \theta,\bm x)).
\end{align*}
The conditional distribution of the latent variable $U|z, y, \bm \theta, \bm x$ is,
\begin{align*}
&p(u|z, y, \bm \theta,\bm x)= \frac{p(z, y, u|\bm \theta, \bm x)}{p(z, y|\bm \theta, \bm x)}
=\frac{[zI(u\geq 0)+(1-z)I(u<0)]p(u, y|\bm \theta, \bm x)}{[(1-\Phi)+z(2\Phi-1)]p(y|\bm \theta, \bm x)}\\
&=p(u|y, \theta, \bm x)\frac{(1-I(u\geq 0))+z(2I(u\geq 0)-1)}{(1-\Phi)+z(2\Phi-1)}.
\end{align*}
In the above equation, $\Phi$ stands for $\Phi(s(y)|\bm \theta, \bm x)$.
We can also write the conditional distribution separately,
\begin{align*}
p(u|y, z=1,\bm \theta, \bm x)&=N\left(u|\bm x'\bm \beta_1+\frac{\rho}{\sigma}(y-\bm x'\bm \beta_2), (1 - \rho^2) \right)\frac{I(u\geq 0)}{\Phi(s(y)|\bm \theta, \bm x)},\\
p(u|y, z=0,\bm \theta, \bm x)&=N\left(u|\bm x'\bm \beta_1+\frac{\rho}{\sigma}(y-\bm x'\bm \beta_2), (1 - \rho^2) \right)\frac{1-I(u\geq 0)}{1-\Phi(s(y)|\bm \theta, \bm x)}.
\end{align*}
Clearly, the latent variable $U$, given the two response variables and the parameters, follows two different truncated normal distributions.

Considering that the outputs of the different experimental runs are independent of each other, the sampling distribution for all the data  $\{\bm x_i, z_i, y_i\}_{i=1}^n$ is
\begin{align*}
&p(\bm z,\bm y|\bm \theta, \bm X)=\prod_{i=1}^n p(z_i,y_i|\bm \theta, \bm x_i)\\\nonumber
  &=\prod_{i=1}^n p(y_i|\bm \theta, \bm x_i)[(1-\Phi(s(y_i)|\bm \theta, \bm x_i))+z_i(2\Phi(s(y_i)|\bm \theta, \bm x_i)-1)]\\
  &=N(\bm y|\bm X\bm \beta_2, \sigma^2\bm I_n)\prod_{z_i=1} \Phi(s(y_i)|\bm \theta, \bm x_i) \prod_{z_i=0} \left(1-\Phi(s(y_i)|\bm \theta, \bm x_i)\right).
\end{align*}
The conditional distribution for $\bm u=(u_1,\ldots, u_n)'$ is
\begin{align*}
&  p(\bm u|\bm y, \bm z, \bm \theta, \bm X)\\\nonumber
&  =N\left(\bm u|\bm X \bm \beta_1+\frac{\rho}{\sigma} (\bm y-\bm X\bm \beta_2), (1 - \rho^2) \bm I_n\right)\prod_{z_i=1}\frac{I(u_i\geq 0)}{\Phi(s(y_i)|\bm \theta, \bm x_i)} \prod_{z_i=0} \frac{1-I(u_i\geq 0)}{1-\Phi(s(y_i)|\bm \theta, \bm x_i)}.
\end{align*}

To obtain the joint posterior distribution $p(\bm \theta|\bm y, \bm z,\bm X)$, we first derive the conditional distribution $p(\bm \beta_1,\bm \beta_2|\bm y, \bm z, \bm u, ,\sigma^2,\rho, \bm X)$.
To simplify the notation, we omit the matrix $\bm X$ in all the conditional side of the distribution, as $\bm X$ is always considered to be known in all Bayesian regression modeling.
We write the previous regression model for $(\bm u, \bm y)$ via the following matrix form
\begin{align*}
\left[
\begin{array}{c}
\bm u \\
\bm y
\end{array}
\right]_{2n \times 1}
= \mathbb{X}\bm \beta + \bm \epsilon, \mbox{where }
\mathbb{X} =\bm I_2\otimes \bm X=\left[
\begin{array}{cc}
\bm X &  \bm 0\\
\bm 0 &  \bm X
\end{array} \right]_{2n \times 2p} \mbox{ and }
\bm \beta=\left[
\begin{array}{c}
\bm \beta_1 \\
\bm \beta_2
 \end{array}
\right]_{2p \times 1}.
\end{align*}
Here $\bm I_2$ is a $2\times 2$ identity matrix and the symbol $\otimes$ stands for the kronecker product between two matrices.
The covariance matrix of the noise $\bm \epsilon$ as well as the vector $(\bm u', \bm y')'$ is
\[\bm \Sigma_{\epsilon} = \cov(\bm \epsilon) = \bm \Sigma \otimes \bm I_n=\left (
\begin{array}{cc}
\bm I_n &  \rho  \sigma \bm I_n\\
\rho  \sigma \bm I_n &  \sigma^2 \bm I_n
\end{array} \right),\]
where $\bm I_n$ is the $n \times n$ identity matrix.
Denote the conjugate prior for $\bm \beta$ as
\begin{align}
\label{eq:beta-prior}
\bm \beta=\left[
\begin{array}{c}
\bm \beta_1 \\
\bm \beta_2
\end{array}
\right] \sim N (\bm 0, \bm \Sigma_0), \mbox{where }
\bm \Sigma_0 = \left[
\begin{array}{cc}
\bm V_1 &  \bm 0\\
\bm 0 &  \bm V_2
\end{array} \right].
\end{align}
Such prior covariance matrix assumes that $\bm \beta_1$ and $\bm \beta_2$ are independent.
Consequently, we have the full-conditional distribution of $\bm \beta$ as following, which does not involve $\bm z$.
\begin{align}\label{eq:beta-post}
\bm \beta=\left.\left[\begin{array}{c}
\bm \beta_1 \\
\bm \beta_2
\end{array}
\right]
\right\vert \bm y, \bm u, \sigma^2, \rho \sim N(\bm \mu_{\bm \beta}, \bm \Sigma_{\bm \beta}),
\end{align}
where the covariance is
\begin{equation}
\label{eq:beta_var}
\bm \Sigma_{\bm \beta} = (\bm \Sigma_0^{-1} + \mathbb{X}' \bm \Sigma_{\epsilon}^{-1} \mathbb{X})^{-1}=(\bm \Sigma_0^{-1}+ \bm \Sigma^{-1}\otimes \bm X'\bm X)^{-1}.
\end{equation}
The inverse matrix $\bm \Sigma^{-1}$ is easily computed by
\begin{align*}
\bm \Sigma^{-1} &= \left [
\begin{array}{cc}
1 & \rho  \sigma \\
\rho  \sigma & \sigma^2
\end{array}
\right]^{-1}
=\frac{1}{(1 - \rho^2)  \sigma^2}\left[\begin{array}{cc} \sigma^2, & -\rho  \sigma,\\ -\rho  \sigma, & 1
\end{array}\right].
\end{align*}
The mean of the full-conditional is
\begin{align}
\label{eq:beta_mu}
\bm \mu_{\bm \beta} &= (\bm \Sigma_0^{-1} + \mathbb{X}' \bm \Sigma_{\epsilon}^{-1} \mathbb{X})^{-1} \mathbb{X}' \bm \Sigma_{\epsilon}^{-1} \left[\begin{array}{c} \bm u \\ \bm y \end{array} \right]
= (\bm \Sigma_0^{-1}+ \bm \Sigma^{-1}\otimes \bm X'\bm X)^{-1} (\bm \Sigma^{-1} \otimes \bm X')\left[\begin{array}{c} \bm u \\ \bm y \end{array} \right]\\\nonumber
& = \frac{1}{(1 - \rho^2) \sigma^2}(\bm \Sigma_0^{-1}+\bm \Sigma^{-1}\otimes \bm X'\bm X)^{-1} \left[
\begin{array}{c}
\sigma^2\bm X'\bm u - \rho  \sigma \bm X'\bm y\\\
-\rho  \sigma \bm X'\bm u + \bm X'\bm y
\end{array}
\right].
\end{align}

In regards of the prior of the parameters $(\sigma^2, \rho)$, we adopt a weekly informative prior for $\sigma^2$ and the uniform prior for $\rho$, $\sigma^2 \sim \textrm{Inv-} \chi^2 (0.001, 0.001)$ and $p(\rho) \sim \textrm{Unif}(-1,1)$. 
Let $\eta_i = u_i- \bm x_i'\bm \beta_1$ and $\varphi_i = y_i- \bm x_i'\bm \beta_2$,
then the posterior distributions of $\sigma^2$ and $\rho$ are easily derived as
\begin{align}\label{eq:sigma}
p(\sigma^2 |\bm y, \bm u, \bm \beta, \rho) &\propto \frac{1}{(\sigma^2)^{\frac{n}{2}}} \exp \{-\frac{1}{2} \sum_{i=1}^n (\eta_i, \varphi_i) \bm \Sigma^{-1} (\eta_i, \varphi_i)' \} \frac{1}{(\sigma^2)^{\frac{2+0.001}{2}}} \exp \{ -\frac{10^{-6}}{2 \sigma^2} \} \\ \nonumber
&\propto \frac{1}{(\sigma^2)^{\frac{n+2+0.001}{2}}} \exp \{ -\frac{1}{2 \sigma^2} [\frac{1}{1 - \rho^2} \sum_{i=1}^n (\sigma^2 \eta_i^2 - 2\rho\sigma\varphi_i\eta_i + \varphi_i^2) + 10^{-6}] \},
\end{align}
and
\begin{align}\label{eq:rho}
p(\rho |\bm y, \bm u, \bm \beta, \sigma) &\propto \frac{1}{(1-\rho^2)^{\frac{n}{2}}}
\exp \{-\frac{1}{2} \sum_{i=1}^n (\eta_i, \varphi_i) \bm \Sigma^{-1} (\eta_i, \varphi_i)' \} \\ \nonumber
&\propto \frac{1}{(1-\rho^2)^{\frac{n}{2}}} \exp\{-\frac{1}{2\sigma^2(1-\rho^2)} \sum_{i=1}^n (\sigma^2 \eta_i^2 - 2\rho\sigma\varphi_i\eta_i + \varphi_i^2)\}.
\end{align}
Since their posteriors are not from any known distributions, the Metropolis-Hasting (MH) algorithm is used to draw the samples of $\sigma^2$ and $\rho$.

\subsection{Leave-one-out Sampling of $\bm u$}\label{subsec:loo}

One might think that the simplest way to sample from $p(\bm \theta|\bm y, \bm z)$ is to use Gibbs sampling method that draws $\bm \theta$ and $\bm u$ iteratively in the follow steps.
\begin{enumerate}
\item $\bm u_j \leftarrow p(\bm u|\bm y, \bm z, \bm \beta_{j-1}, \sigma^2_{j-1}, \rho_{j-1})$,
\item $\bm \beta_{j} \leftarrow p(\bm \beta |\bm y, \bm u_{j}, \sigma^2_{j-1}, \rho_{j-1})$,
\item $\sigma^2_{j} \leftarrow p(\sigma^2 |\bm y, \bm u_j, \bm \beta_{j}, \rho_{j-1})$
\item $\rho_{j} \leftarrow p(\rho |\bm y, \bm u_j, \bm \beta_{j}, \sigma^2_j)$.
\end{enumerate}
However, as discussed in \cite{holmes2006bayesian}, a potential problem lurks in the strong posterior correlation between $\bm \beta_1$ and $\bm u$, as assumed in the model $\bm u|\bm \theta \sim N(\bm X\bm \beta_1, \bm I_n)$.
This strong correlation would cause slow mixing in the MCMC chain and thus leads to large computation.
Instead, we follow the approach suggested by \cite{holmes2006bayesian} and update $\bm \beta$ and $\bm u$ jointly by making the factorization
\[
p(\bm \beta, \bm u|\bm y, \bm z, \sigma^2, \rho)=p(\bm u|\bm y, \bm z, \sigma^2, \rho)p(\bm \beta|\bm y, \bm u, \sigma^2, \rho).
\]
The distribution $p(\bm \beta|\bm y, \bm u, \sigma^2, \rho)$ is the normal distribution in \eqref{eq:beta-post}.
The distribution $p(\bm u|\bm y, \bm z, \sigma^2, \rho)$ can be obtained by integrating $p(\bm \beta)p(\bm u|\bm \beta,\bm y,\bm z,\sigma^2, \rho)$ with respect to $\bm \beta$.
Given the prior of $\bm \beta$ in \eqref{eq:beta-prior}, we can obtain
\[
\bm u|\bm y, \bm z, \sigma^2, \rho \sim \bm N\left(\frac{\rho}{\sigma} \bm y, (1 - \rho^2) \bm I_n + \bm X \bm V_1\bm X'+  \frac{\rho^2}{\sigma^2}\bm X\bm V_2\bm X'\right)Ind(\bm y, \bm z, \bm u),
\]
where $Ind(\bm y, \bm z, \bm u)$ is an indicator function that truncates the multivariate normal distribution into the appropriate region.
It is well-known that directly sampling from a truncated multivariate normal distribution is difficult, as pointed out by \cite{holmes2006bayesian}.
Hence, we use a more straightforward Gibbs sampling method,
\[
u_i|\bm u_{-i}, \bm y, z_i, \sigma^2, \rho \sim
\left\{
\begin{array}{ll}
N(m_i, v_i)I(u_i\geq 0), & \textrm{ if } z_i=1,\\
N(m_i, v_i)I(u_i<0), & \textrm{ if }z_i=0,
\end{array}
\right.
\]
where $\bm u_{-i}$ denotes all the latent variables $\bm u$ without $u_i$.
The mean $m_i$ and variance $v_i$ for $i=1,\ldots, n$ are obtained from the leave-one-out marginal predictive distributions, and its derivation is in Appendix A in the supplement document.
\begin{align*}
m_i& = \frac{\rho}{\sigma} y_i + \left[\bm x_i',-\frac{\rho}{\sigma} \bm x_i'\right]\bm \mu_{\bm \beta,-i},\quad
v_i=\left[\bm x_i',-\frac{\rho}{\sigma}\bm x_i'\right]\bm \Sigma_{\bm \beta,-i}\left[\begin{array}{c} \bm x_i, \\ -\frac{\rho}{\sigma}\bm x_i \end{array}\right] + (1 - \rho^2).
\end{align*}
The notations $\bm \mu_{\bm \beta,-i}$ and $\bm \Sigma_{\bm \beta,-i}$ are the mean and covariance matrix of the distribution $\bm \beta|\bm u_{-i}, \bm y, \sigma^2, \rho$.
Since these two need to be calculated frequently, we have derived a shortcut formula to facilitate the computation in Appendix B in the supplement document.

\section{MCMC sampling} \label{sec:MCMC}
In this section, we specify the prior distributions for the parameter $\bm \beta$ as well as hyperprior distributions for the hyperparameters $r_1$, $r_2$, $\tau_1^2$, $\tau_2^2$.
Then the corresponding posteriors of these parameters are obtained.
The Gibbs sampling algorithm is laid out to sample the posterior distributions.

\subsection{Prior and hyperprior distributions}

The marginal prior components for $\bm \beta_1$ and $\bm \beta_2$ are
\begin{equation}\label{eq:priors}
\bm \beta_i\sim N({\bf 0}, \tau_i^2\bm R_i) \textrm{ for }i=1, 2.
\end{equation}
The correlation matrices in \eqref{eq:priors} in the marginal prior components for $\bm\beta_1$ and $\bm \beta_2$ are assumed to be diagonal, which means that the coefficients are independent of each other.
This assumption is reasonable if we use the orthogonal polynomial basis of $\bm x$, consisting of the intercept, the linear effects, the quadratic effects, and the interactions, etc., up to a user-specified order.
If the controllable variable settings are from a full factorial design or an orthogonal design, we can achieve full or near orthogonality between the bases.
For the bases involving covariates, it is not likely to achieve full- or near-orthogonality.
But we still assume independence for simplicity and leave the data to correct it in the posterior distribution.
Let $\bm R_i=\diag\{1, r_i, \ldots, r_i, r_i^2, \ldots, r_i^2, \ldots\}$ for $i=1,2,3$, where $r_i\in (0,1)$ is a user-specified tuning parameter.
The power index of $r_i$ is the same as the order of the corresponding polynomial term.
For example, if the polynomial regression terms of $\bm x \in \mathbb{R}^2$ is a full quadratic model and contains the term $\{1,x_1,x_2,x_1^2,x_2^2,x_1x_2\}$, the corresponding prior correlation matrix should specified as $\bm R=\diag\{1,r,r,r^2,r^2,r^2\}$.
In this way, the prior variance of the effect is decreasing exponentially as the order of effect increases, following the hierarchy ordering principle defined in \cite{wu2011experiments}.
The hierarchy ordering principle can reduce the size of the model and avoid including higher-order and less significant model terms.
Such prior distribution was firstly proposed by \cite{joseph2006bayesian}, and later used by \cite{kang2009bayesian, ai2009bayesian}.
It is also used in another Bayesian QQ model by \cite{kang2018bayesian}.

Additionally, we use the hyperprior distributions for the hyperparameters $\tau_1^2, \tau_2^2 \sim _{iid} \textrm{Inv-}\chi^2(\nu, \delta^2)$ and $r_1, r_2 \sim_{iid} \textrm{Beta}(a, b)$, where $\textrm{Inv-}\chi^2(\nu, \delta^2)$ stands for the scaled inverse-chi-square distribution with $\nu$ degrees of freedom and scale $\delta^2$.
Beta distribution is a reasonable prior for $r_i$ since $r_i \in (0,1)$.
Then it is not difficult to derive the posterior distributions for $r_1, r_2, \tau_1^2$ and $\tau_2^2$ listed below
\begin{align}
\label{eq:tau1}
&\tau_{1}^{2} | \mbox{rest parameters}, \bm y, \bm z \sim \textrm{Inv-}\chi^2(\nu + p,~\frac{1}{\nu + p}[\bm \beta_1' \bm R_{1}^{-1} \bm \beta_1 + \nu \delta^2]), \\
\label{eq:tau2}
&\tau_{2}^{2} | \mbox{rest parameters}, \bm y, \bm z \sim \textrm{Inv-}\chi^2(\nu + p,~\frac{1}{\nu + p}[\bm \beta_2' \bm R_{2}^{-1} \bm \beta_2 + \nu \delta^2]), 
\end{align}
\begin{align}
\label{eq:r1}
& p(r_{1} | \mbox{rest parameters}, \bm y, \bm z) \propto |\bm R_{1}|^{-\frac{1}{2}}~\mbox{exp} \{-\frac{1}{2 \tau_{1}^{2}} \bm \beta_1' \bm R_{1}^{-1} \bm \beta_1\}~r_{1}^{a-1} (1-r_{1})^{b-1}, \\
\label{eq:r2}
&p(r_{2} | \mbox{rest parameters},\bm y, \bm z) \propto |\bm R_{2}|^{-\frac{1}{2}}~\mbox{exp} \{-\frac{1}{2 \tau_{2}^{2}} \bm \beta_2' \bm R_{2}^{-1} \bm \beta_2\}~r_{2}^{a-1} (1-r_{2})^{b-1}.
\end{align}
We directly sample $\tau_{1}^{2}$ and $\tau_{2}^{2}$ from their respective scaled inverse-chi-square distributions, and we use Metropolis-Hastings (MH) algorithm is applied to sample $r_1$ and $r_2$ from \eqref{eq:r1} and \eqref{eq:r2}.

\subsection{Gibbs sampling algorithm}

We use the following Gibbs sampling algorithm to generate the posterior distributions for the (hyper)parameters and the latent variable.
\begin{enumerate}
\item [Step 0] Set up the initial values for the parameters and the latent variable. Set the counter $j=0$. For the counter $j=1,2, \ldots, B$.
\item [Step 1]Sample $\bm u_{j}$ from $p(\bm u|\bm y, \bm z, \sigma^2_{j-1}, \rho_{j-1})$ by drawing $u_{i,j}$ from the leave-one-out marginal distribution $p(u_i|\bm u_{-i,j-1}, \bm y, z_i, \sigma^2_{j-1}, \rho_{j-1})$ for $i=1,\ldots, n$.
\item [Step 2] Sample $\bm \beta_{j}$ from $p(\bm \beta|\bm y, \bm u_{j}, \sigma^2_{j-1}, \rho_{j-1})$ according to \eqref{eq:beta_var} and \eqref{eq:beta_mu}.
\item [Step 3] Sample $\sigma^2_{j}$ and $\rho_{j}$ from \eqref{eq:sigma} and \eqref{eq:rho} by the MH algorithm.
\item [Step 4] Sample $\tau_{1, j}^{2}$ and $\tau_{2, j}^{2}$ from \eqref{eq:tau1} and \eqref{eq:tau2}.
\item [Step 5] Sample $r_{1,j}$ and $r_{2,j}$ by the MH algorithm from the distributions \eqref{eq:r1} and \eqref{eq:r2}.
\item [Step 6] Do Step 1--Step 5 until the MCMC chain converges.
\end{enumerate}

The initial values of $\bm \beta_2$ are set to be the least square estimate from $\bm y=\bm X\bm \beta_2+\bm \epsilon_2$, and the initial $\sigma^2$ value is the mean squared error of the linear regression model.
The initial values of $\bm \beta_1$ are the MLE of the probit regression of $\bm z$ with the same model matrix $\bm X$.
The estimated link function values of the probit regression can be the initial values of $\bm u$.
The initial value of $\rho$ is calculated from the sample correlation between $\bm u$ and $\bm y$.

In Step 1, given the current $\bm \Sigma_{\bm \beta}$ and $\bm \mu_{\bm \beta}$, we use the short-cut formula in Appendix B (in supplement document) to calculate $\bm \Sigma_{\bm \beta,-i}$ and $\bm \mu_{\bm \beta,-i}$.
After each $u_i$ is updated, we need to update the vector of $\bm u_{j-1}$ to be $(u_{1,j}, \ldots, u_{i,j}, u_{i+1,j-1},\ldots, u_{n,j-1})$.
The covariance $\bm\Sigma_{\bm \beta}$ remains the same for all $i$, but $\bm \mu_{\bm \beta,-i}$ needs to be updated using $(u_{1,j}, \ldots, u_{i,j}, u_{i+1,j-1},\ldots, u_{n,j-1})$.

\section{Numerical study}\label{sec:simulation}
In this section, the performance of the proposed model is examined and compared with two approaches SM(F) and SM(B), where the qualitative variable $Z$ and quantitative variable $Y$ are modeled separately.
Hence, both SM(F) and SM(B) ignore the association between variables $Z$ and $Y$.
SM(F) employs a logistic model for the variable $Z$, and a linear regression model to fit $Y$.
The LASSO regularization is applied for both logistic and linear regression models to select the significant variables.
SM(B) denotes the separate modeling of $Z$ using probit regression and of $Y$ using linear regression under the Bayesian framework.
SM(B) sets the marginal normal priors for the parameters in both linear and probit models.

Since the parameter $\rho$ reflects the strength and direction of the relationship between the value of $Y$ and the probability $Z = 1$, we consider three different cases:
(1) $\rho = 0$; (2) $\rho = 0.85$; (3) $\rho = -0.5$.
In each case, we generate $n=100$ training data points and $n=100$ testing data points based on the models \eqref{eq: model-1} and \eqref{eq: model-2}.
All data are independently and identically distributed from normal with mean $\bm 0$ and covariance matrix $\bm \Sigma_x = (\sigma_{ij})_{p \times p}$ with $\sigma_{ij} = 0.5^{|i-j|}$.
The variance $\sigma^2$ in the model \eqref{eq: model-2} is set to be 2.
To further examine the performance of the proposed model,
we consider the settings of different model size $p \in \{10, 30\}$ and proportion of sparsity $s\in \{20\%, 50\% \}$, where the value of $s$ represents the proportion of nonzero entries
in the parameter vector $\bm \beta_1$ and $\bm \beta_2$.
Overall, the full combinations have $3 \times 2 \times 2 = 12$ settings.

For the true values of $\bm \beta_1$ and $\bm \beta_2$,
we first randomly place zeroes in $\bm \beta_1$ and $\bm \beta_2$.
Then the values of non-zeroes are generated from $N(3, 1)$ independently,
with positive signs and negatives signs randomly assigned to the non-zeroes elements of $\bm \beta_1$ and $\bm \beta_2$.
To evaluate the estimation accuracy of the estimates obtained from each approach with respect to $\bm \beta_1$ and $\bm \beta_2$, we use the loss measures as follows:
\begin{align*}
L_2(\hat{\bm \beta}_1) = || \hat{\bm \beta}_1 - \bm \beta_1 ||_2^2~~~ \mbox{and} ~~~
L_2(\hat{\bm \beta}_2) = || \hat{\bm \beta}_2 - \bm \beta_2 ||_2^2,
\end{align*}
where $||\cdot||_2$ denotes the vector $L_2$ norm.
Additionally, to gauge the performance of variable selection for $\bm \beta = (\bm \beta_1', \bm \beta_2')'$, false positive (FP) and false negative (FN) cases are considered.
We say an FP occurs if a nonsignificant predictor in the true model is incorrectly identified as a significant one.
Similarly, an FN occurs if a significant predictor in the true model is incorrectly estimated as a nonsignificant one.
We report $FSL$ = FP + FN, the total number of FP and FN cases, as the performance measure of variable selection.
In the SM(F) method, the significant predictors are selected by the LASSO.
For the proposed model and SM(B), the variable selection is conducted based on the 95\%
credible intervals constructed from the MCMC samples after the burn-in period.
Furthermore, we evaluate the model's prediction capacity using the root-mean-square error
$RMSE = \sqrt{\frac{1}{n} \sum_{i=1}^n (y_i - \hat{y}_i)^2}$ for the quantitative variable
$Y$, where $\hat{y}_i$ is the predicted value for $y_i$ in the testing data set.
We define $ME = \frac{1}{n} \sum_{i=1}^n I_{(z_i \neq \hat{z}_i)}$ to measure the model's prediction performance on the qualitative variable $Z$, where $I_{(\cdot)}$ stands for the indicator function and $\hat{z}_i$ is the predicted value for $z_i$.
For the proposed model, we use $(\nu, \delta^2, a, b)= (2,2,0.1,0.1)$
and initial values $(\tau_{1, 0}^{2}, \tau_{2, 0}^{2}, r_{1,0}, r_{2,0})= (0.5, 0.5, 0.3, 0.3)$.
We set the length of the MCMC chain to be 10000 iterations with the first 1000 as the burn-in period.
Tables \ref{table:p10} and \ref{table:p30} report the simulation results for each loss measure of estimates obtained from each approach over 50 replicates.
Only the proposed approach (BLQQ column) shows the average and standard error (in the parenthesis) of the 50 replicates of the estimated $\hat{\rho}$.

From Tables \ref{table:p10} - \ref{table:p30}, we observe the following results.
\begin{itemize}
\item In the case of $\rho = 0$, the proposed method is comparable to SM(F) and slight better than SM(B) in terms of $RMSE$. Regarding the loss $ME$, the proposed method shows a better performance when $p = 10$ and a comparable, sometimes even worse performance when $p = 30$. The proposed method is always inferior to SM(B) with respect to $FSL$.
Additionally, the proposed method performs the best under $L_2(\hat{\bm \beta}_2)$.
However for $L_2(\hat{\bm \beta}_1)$, the proposed method is worse than SM(B) when $p = 10$ and better than SM(B) in the case of $p = 30$.
Overall, the proposed method performs comparably when $\rho = 0$.
This is expected since there is no association between the variables $Y$ and $Z$. Hence, the proposed joint model does not show its advantages.
\item  When $\rho = 0.85$, the proposed method remarkably outperforms the other two approaches, since SM(F) and SM(B) ignore the dependency between variables $Y$ and $Z$ in this case.
Specifically, the proposed method gives superior performance over SM(F) regarding every criterion, especially in terms of $FSL$ and $L_2(\hat{\bm \beta}_1)$.
Compared with SM(B), although the proposed method is comparable or even inferior under $FSL$ when the model is sparse as $s = 0.2$, it is better when the true model becomes denser as $s = 0.5$. For other comparison criteria, the proposed method greatly outperforms SM(B).
The results from this case demonstrate the advantages of the proposed joint model over the separate models.
\item When the variables $Y$ and $P(Z = 1)$ are negatively correlated as $\rho = -0.5$, the conclusions are very similar to those for $\rho = 0.85$.
The proposed method consistently outperforms SM(F) and SM(B) because of the dependency between $Y$ and $Z$.
\item The proposed method can give an estimate of $\rho$, while the other two approaches cannot. This correlation indicates both the strength and direction of the association between $Y$ and the probability of $Z = 1$.
Hence, the estimated $\hat{\rho}$ provides us with more insight to understand data.
\end{itemize}

\begin{figure}
\centering
\scalebox{0.5}[0.5]{\includegraphics{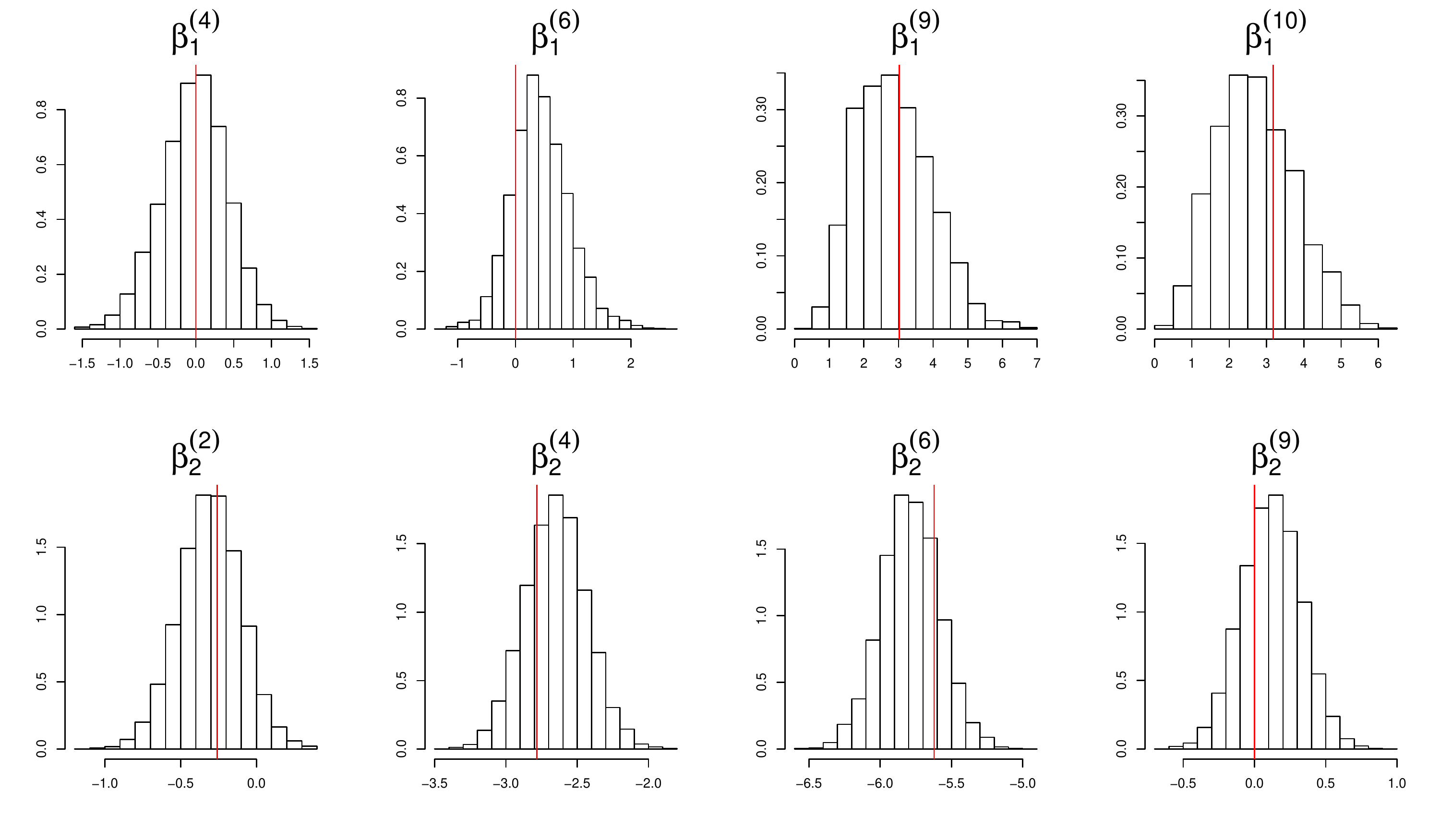}}
\caption{Histograms for the selected parameters of one replicate from $\rho = -0.5$ when $p = 10$ and $s = 0.5$.}\label{BLQQ:hist_sim}
\end{figure}

For illustration, based on a single simulation, Figure \ref{BLQQ:hist_sim} displays the histograms for the posterior samples of some parameters after the burn-in period with their true values indicated by the red vertical lines.
Such histogram and posterior distributions can be used for inferences.

\section{Birth Records Case Study}\label{sec:app}
In this section, we apply the proposed method to evaluate its utility in evaluating factors associated with preterm birth and birth weight, as described in the Introduction.  The birth record dataset was acquired from the Virginia Department of Health via a Data Sharing Agreement and this application is approved by the Virginia Department of Health Institutional Review Board (IRB) (Protocol \#40221) and Virginia Tech IRB (Protocol \#16-898).  The full dataset includes over three million observations for more than two decades.
Only a subset of the data was used for this study with a total of $1,000$ observations.
In the original dataset, the binary outcome variable ``preterm birth'' is extremely skewed as preterm births, in general, account for less than $10\%$ of all live births.
We choose a random sample of $n=1,000$ that is more balanced with an equal number of preterm births and non-preterm births.
This balancing is done for computational reasons.
Further enhancements to the model to handle unbalanced data are feasible due to the Bayesian specification.

There are $9$ covariates contained in this dataset, along with the two outcome variables of interest ``preterm birth'' or PTB, which is dichotomous, and ``Birth Weight'', which is continuous (measured in grams).
The covariates include the age of mother, day of birth, day of the week (previous research has shown seasonal as well as weekly patterns for preterm birth. e.g.,
\citep{Darrow2009, Palmer2015}, parity number (whether this is the first pregnancy carried to 24 weeks gestation or not), college education of mother (a proxy for socio-economic status of the mother), etc.
The more detailed description is given in Table \ref{tab:case}.
Intuitively, the two outcome variables are negatively correlated as children who experience preterm births are also more likely to have lower birth weight.

First, we compare the different modeling approaches.
The number of MCMC iterations is set to be 10000 with the burn-in period of 2000.
We use $(\nu, \delta^2, a, b)= (2,2,0.1,0.1)$ and initial values $(\tau_{1, 0}^{2}, \tau_{2, 0}^{2}, r_{1,0}, r_{2,0})= (1.5, 3, 0.3, 0.3)$ for the proposed Bayesian model.
To evaluate its performance, the whole data set is randomly split into a training set with 100 observations and a testing set with 900 observations.
Such partitions are repeated 50 times.
For each random split, we fit the training data by SM(F), SM(B), the Bayesian Hierarchical QQ Model by \cite{kang2018bayesian} (BHQQ for short) and the proposed Bayesian Latent QQ model (BLQQ) and we make predictions on the testing data.

\begin{figure}
\begin{center}
\scalebox{0.5}[0.45]{\includegraphics{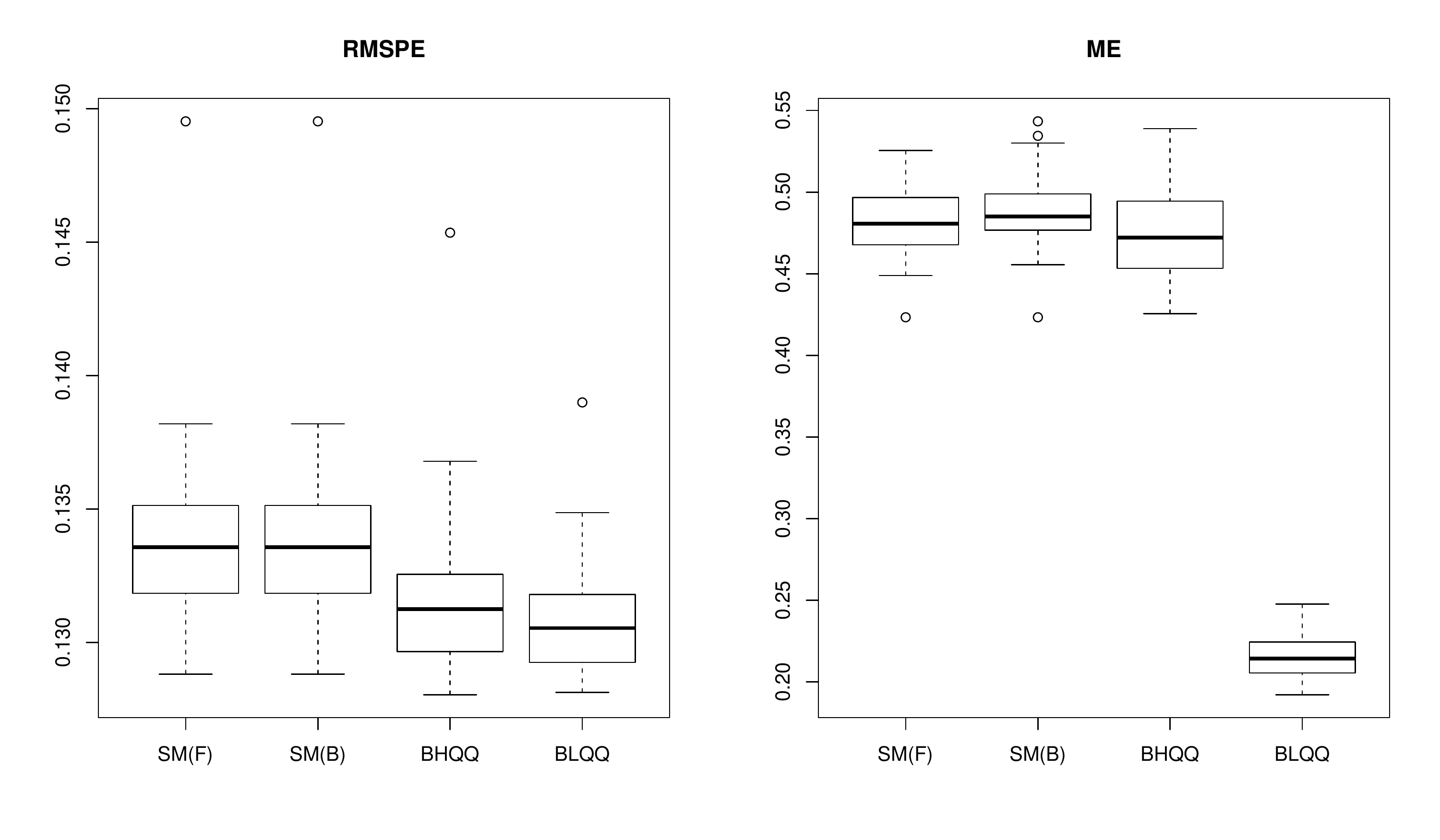}}
\caption{Boxplots of RMSPE and mis-classification error for preterm birth data for each approach.}
\label{Figure:case2pred}
\end{center}
\end{figure}

Figure \ref{Figure:case2pred} shows the root mean square prediction error (RMSPE) and misclassification error (ME) for each method.
The separate models SM(F) and SM(B) perform similarly to each other while the proposed method shows better performance than both of them because of the dependency of two outcome variables.
The proposed model gives a significantly lower ME, indicating that it can distinguish the preterm births from non-preterm births much more accurately.
The proposed model is also better in predicting the birth weight as shown in the boxplot of RMSPE.
Besides, the proposed method can account for the correlation between birth weight and the probability of PTB.
The average of the estimated correlation over 50 splits is -0.772 with a standard error 0.063.
We also note that for each split of the data set, the estimated correlation is negative.
It means the smaller value of the birth weight variable, the more likely the corresponding birth is preterm.
Note that the latest method BHQQ is comparable with the proposed method in terms of prediction accuracy for the continuous outcome, but is much worse regarding ME.
This is expected as we have explained in the Introduction.
BHQQ uses the marginal logistic regression model for the binary outcome and thus cannot improve the prediction accuracy for the binary outcome.

\begin{figure}
\begin{center}
\scalebox{0.5}[0.5]{\includegraphics{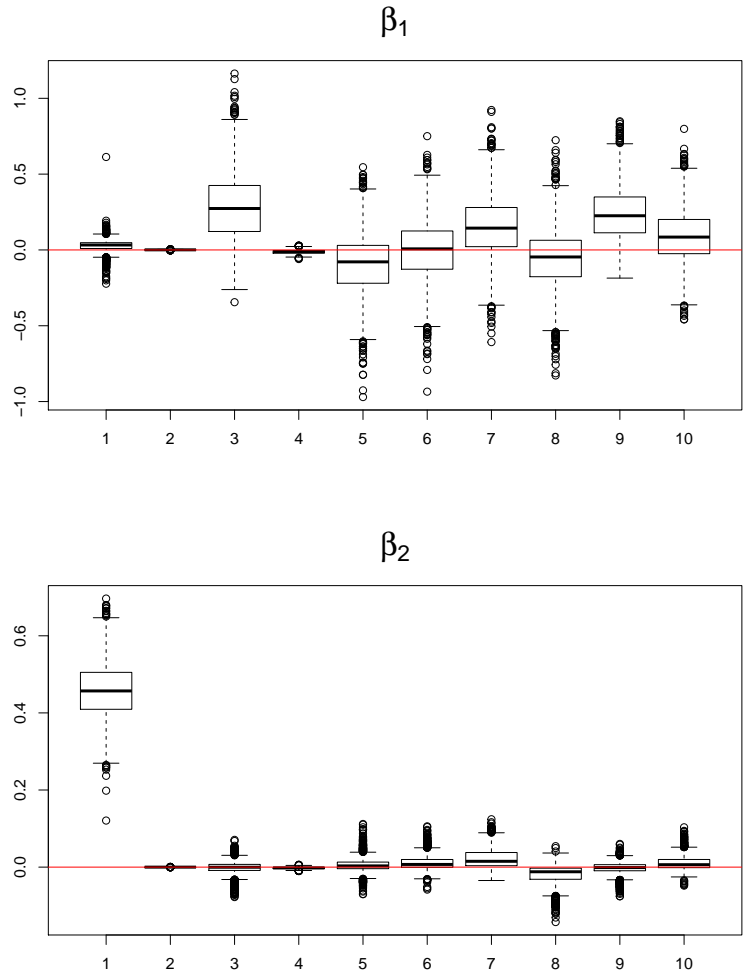}}
\caption{Regression coefficient distributions for the explanatory variables (1 indicates the regression constant) for the quantitative and qualitative responses across 50 replications.}
\label{Fig:betas}
\end{center}
\end{figure}

Next, we investigate the analysis results based on one random split of the training and testing data sets.
There are 500 observations with PTB = 1 and 500 observations with PTB = 0 in the testing set. (a total of $1,000$ observations.
The estimate of the correlation is -0.85. We verified from trace plots that Gibbs sampling iterations converge and using the ACF plots that the autocorrelation dies off. We omit these plots in the paper. We show boxplots of the regression coefficients in Figure \ref{Fig:betas} across the 50 replications (the X-axis is numbered from $1$ to $10$ to indicate regression constant and the slopes corresponding to the $9$ explanatory variables). The first subplot corresponds to the regression coefficients for the qualitative response (preterm birth) and the second subplot corresponds to the regression coefficients for the quantitative response (birth weight). Given the complex biological and physiological causes of preterm births and birth weights of children, it is not surprising that the regression coefficients are not statistically significant at the default $0.05$ level.

\section{Discussion}\label{sec:end}
In this article, we propose a Bayesian latent variable model to jointly fit data with qualitative and quantitative (QQ) outcomes.
The work is motivated by a birth records study involving two responses: birth weight (quantitative variable) and preterm birth (qualitative variable).
The proposed model uses a latent variable to link the quantitative and qualitative variables, improving the prediction accuracy for both variables, while some existing works without using a latent variable fit one variable conditional on the other variable, hence improving the prediction accuracy for only one variable.
Moreover, the proposed model can capture the correlation between the quantitative variable and the latent variable, which is an indicator of the dependency strength for the quantitative and qualitative variables.
Besides, the proposed Bayesian framework is more convenient to provide statistical inference for the parameters than the frequentist analysis based on the asymptotic distribution of the estimator, which is complicated and difficult to derive.
The merits of the proposed Bayesian latent variable model is demonstrated by the numerical study and a birth records data set.

\section*{Acknowledgment}

This research was partially supported by National Science Foundation grant DMS-1916467. 

\section*{Appendix A}

\noindent The leave-one-out predictive distribution for $u_i|\bm u_{-i}, \bm y, \bm z,\sigma^2, \rho$ can be obtained through
\[p(u_i|\bm u_{-i}, \bm y, \bm z, \sigma^2, \rho)=\int p(u_i|y_i, z_i, \bm \beta,\sigma^2, \rho) p(\bm \beta|\bm u_{-i}, \bm y, \bm z,\sigma^2, \rho)d\bm \beta,\]
where $p(\bm \beta|\bm u_{-i}, \bm y, \bm z, \sigma^2, \rho)$ can be derived in the same way as we did for \eqref{eq:beta-post}.
The sampling distribution of $(\bm u_{-i}, \bm y)$ is directly obtain as
\[
\left.\left[
\begin{array}{c}
\bm u_{-i} \\
\bm y
\end{array}
\right]\right\vert\bm \theta \sim N\left(\mathbb{X}_{-i}\bm \beta,\bm \Sigma_{\epsilon,-i}\right),
\]
where $\mathbb{X}_{-i}$ is the $\mathbb{X}$ matrix with its $i$th row removed, i.e.,
\[
\mathbb{X}_{-i}=\left[
\begin{array}{cc}
\bm X_{-i}, & {\bf 0}_{(n-1)\times p}\\
{\bf 0}_{n\times p}, & \bm X
\end{array}
\right].
\]
Here $\bm X_{-i}$ is $\bm X$ with its $i$th row removed.
The covariance matrix $\bm \Sigma_{\epsilon,-i}$ is $\bm \Sigma_{\epsilon}$ without the $i$th row and $i$th column removed.
For convenience, permute the rows and columns of $\bm \Sigma_{\epsilon}$ so that the $i$th row and column are the last,
\[
\bm \Sigma_{\epsilon}=\left[
\begin{array}{cc}
\bm\Sigma_{\epsilon,-i}, & \bm l \\
\bm l', & 1
\end{array}
\right],
\]
where $\bm l=\left[{\bf 0}_{1\times (n-1)},0,\ldots,0, \rho \sigma,0,\ldots,0\right]$.
So all elements of $\bm l$ are zeroes except the $(n-1)+i$th element is $\rho \sigma$.
Since the prior of $\bm \beta$ is $N({\bf 0},\bm \Sigma_0)$, the full-conditional distribution of $\bm \beta$ conditioned on $(\bm u_{-i},\bm y)$ is
\[
\bm \beta|\bm u_{-i}, \bm y, \sigma^2, \rho \sim N(\bm \mu_{\bm \beta,-i}, \bm \Sigma_{\bm \beta,-i}).
\]
Through directly calculation,
\begin{align*}
\bm \Sigma_{\bm \beta,-i}&=\left(\bm \Sigma_0^{-1}+\mathbb{X}_{-i}'(\bm \Sigma_{\epsilon,-i})^{-1}\mathbb{X}_{-i}\right)^{-1},\\
\bm \mu_{\bm \beta,-i}&=\bm \Sigma_{\bm \beta,-i}\mathbb{X}_{-i}'(\bm \Sigma_{\epsilon,-i})^{-1}\left[\begin{array}{c}\bm u_{-i} \\ \bm y \end{array}\right].
\end{align*}
Previously, we have shown that
\[
u_i|y_i,z_i,\bm \theta \sim \left\{
\begin{array}{ll}
N\left(\bm x_i'\bm \beta_1+\frac{\rho}{\sigma}(y_i-\bm x_i'\bm \beta_2), (1 - \rho^2) \right)\frac{I(u_i\geq 0)}{\Phi(s(y_i)|\bm \theta)}, &\quad\textrm{ if }z_i=1,\\
N\left(\bm x_i'\bm \beta_1+\frac{\rho}{\sigma}(y_i-\bm x_i'\bm \beta_2), (1 - \rho^2) \right)\frac{I(u_i < 0)}{1-\Phi(s(y_i)|\bm \theta)}, &\quad \textrm{ if }z_i=0,
\end{array}
\right.
\]
and
\[
u_i|y_i,\bm \theta \sim N\left(\bm x_i'\bm \beta_1+\frac{\rho}{\sigma}(y_i-\bm x_i'\bm \beta_2), (1 - \rho^2) \right).
\]
We first derive the distribution for $u_i|\bm u_{-i}, \bm y, \sigma^2, \rho$, which should also be a normal distribution.
Its mean and variance are
\begin{align*}
m_i&=E(u_i|\bm u_{-i},\bm y, \sigma^2, \rho)\\
&=E_{\bm \beta}\left(E_{u_i}\left(u_i|y_i, \bm \beta,\sigma^2, \rho\right)|\bm u_{-i},\bm y, \sigma^2, \rho\right)\\
&=E_{\bm \beta}\left(\bm x_i'\bm \beta_1+\frac{\rho}{\sigma}(y_i-\bm x_i'\bm \beta_2)|\bm u_{-i},\bm y, \sigma^2, \rho\right)\\
&=\frac{\rho}{\sigma}y_i+\left[\bm x_i',-\frac{\rho}{\sigma}\bm x_i'\right]\bm \mu_{\bm \beta,-i}
\end{align*}
and
\begin{align*}
v_i&=\var\left(u_i|\bm u_{-i},\bm y, \sigma^2, \rho\right)\\
&=\var_{\bm \beta}\left(E_{u_i}\left(u_i|y_i,\bm \beta,\sigma^2, \rho \right)|\bm u_{-i}, \bm y, \sigma^2, \rho\right)+E_{\bm \beta}\left(\var\left(u_i|y_i,\bm \beta,\sigma^2, \rho\right)|\bm u_{-i}, \bm y, \sigma^2, \rho\right)\\
&=\var_{\bm \beta}\left(\bm x_i'\bm \beta_1+\frac{\rho}{\sigma}(y_i-\bm x_i'\bm \beta_2)|\bm u_{-i}, \bm y, \sigma^2, \rho\right)+
E_{\bm \beta}\left((1 - \rho^2) 1|\bm u_{-i}, \bm y, \sigma^2, \rho\right)\\
&=\left[\bm x_i',-\frac{\rho}{\sigma}\bm x_i'\right]\bm \Sigma_{\bm \beta,-i}\left[\begin{array}{c} \bm x_i, \\-\frac{\rho}{\sigma}\bm x_i \end{array}\right]+(1 - \rho^2).
\end{align*}
Therefore, the leave-one-out distribution for $u_i|\bm u_{-i}, \bm y, \sigma^2, \rho$ is $N(m_i, v_i)$.
Adding $\bm z$, we obtain
\[p(u_i|\bm u_{-i}, \bm y, z_i,\sigma^2, \rho)\propto \left\{
\begin{array}{cc}
N(u_i|m_i,v_i)I(u_i\geq 0),\textrm{ if }z_i=1,\\
N(u_i|m_i,v_i)I(u_i< 0),\textrm{ if }z_i=0.
\end{array}
\right.
\]

\section*{Appendix B}

Since we have to compute $\bm \Sigma_{\epsilon,-i}$ and $\bm \mu_{\bm \beta,-i}$ for each $u_i$ in each sampling of $\bm u$, thus it is necessary to find a quick way to compute both.
Consider we have already computed $(\bm \Sigma_{\epsilon})^{-1}$ and $\bm \Sigma_{\bm \beta}$.
It can be shown that
\[
\bm \Sigma_{\epsilon}^{-1}=\left[
\begin{array}{cc}
(\bm \Sigma_{\epsilon,-i})^{-1}+c\left((\bm \Sigma_{\epsilon,-i})^{-1}\bm l \bm l'(\bm \Sigma_{\epsilon,-i})^{-1}\right), & -c(\bm \Sigma_{\epsilon,-i})^{-1}\bm l\\
-c\bm l'(\bm \Sigma_{\epsilon,-i})^{-1}, & c
\end{array}
\right],
\]
where $c$ is the diagonal entry of $\bm \Sigma_{\epsilon}^{-1}$ and $c=\left(\sigma_1^2-\bm l'\bm \Sigma_{\epsilon,-i}^{-1}\bm l\right)^{-1}$.
So
\[
(\bm \Sigma_{\epsilon,-i})^{-1}=\left(\bm \Sigma_{\epsilon}^{-1}\right)_{-i,-i}-c\left((\bm \Sigma_{\epsilon,-i})^{-1}\bm l \bm l'(\bm \Sigma_{\epsilon,-i})^{-1}\right)=\left(\bm \Sigma_{\epsilon}^{-1}\right)_{-i,-i}-c^{-1}\left(\bm \Sigma_{\epsilon}^{-1}\right)_{-i,i}\left(\bm \Sigma_{\epsilon}^{-1}\right)_{i,-i}.
\]
Here $\left(\bm \Sigma_{\epsilon}^{-1}\right)_{-i, i}$ is the $i$th column of matrix $\bm \Sigma_{\epsilon}^{-1}$ without the $i$th diagonal entry $\left(\bm \Sigma_{\epsilon}^{-1}\right)_{ii}$ and $(\bm \Sigma_{\epsilon}^{-1})_{i,-i}$ is the $i$th row of $\bm\Sigma_{\epsilon}^{-1}$ without the $i$th diagonal entry, and $(\bm \Sigma_{\epsilon})^{-1}_{-i,-i}$ is the matrix $\bm \Sigma_{\epsilon}^{-1}$ with the $i$th row and $i$th column removed.

Define
\[
\bm b=\mathbb{X}_i-\mathbb{X}_{-i}'(\bm \Sigma_{\epsilon,-i})^{-1}\bm l=\mathbb{X}_i+c^{-1}\mathbb{X}_{-i}'\left(\bm \Sigma_{\epsilon}^{-1}\right)_{-i,i}.
\]
The column vector $\mathbb{X}_i$ is the transpose of the $i$th row of $\mathbb{X}$.
We also can compute following.
\begin{align*}
\mathbb{X}'\bm \Sigma_{\epsilon}^{-1}\mathbb{X}&=\mathbb{X}_{-i}'(\bm \Sigma_{\epsilon,-i})^{-1}\mathbb{X}_{-i}+ c \bm b \bm b',\\
\bm \Sigma_{\bm \beta}&=\left(\bm \Sigma_0^{-1}+\mathbb{X}'\bm \Sigma_{\epsilon}^{-1}\mathbb{X}\right)^{-1}\\
&=\left(\bm \Sigma_0^{-1}+\mathbb{X}_{-i}'(\bm \Sigma_{\epsilon,-i})^{-1}\mathbb{X}_{-i}+ c \bm b \bm b'\right)^{-1}\\
&=\left((\bm \Sigma_{\bm \beta,-i})^{-1}+ c \bm b \bm b'\right)^{-1}.
\end{align*}
Thus,
\begin{align*}
\bm \Sigma_{\bm \beta,-i}&=\left((\bm \Sigma_{\bm \beta})^{-1}-c\bm b \bm b'\right)^{-1}\\
&=\bm \Sigma_{\bm \beta}+\frac{c}{1-c\bm b'\bm \Sigma_{\bm \beta}\bm b}\bm \Sigma_{\bm \beta}\bm b\bm b'\bm \Sigma_{\bm \beta}.
\end{align*}
The vector $\bm \Sigma_{\bm \beta}\bm b$ can be obtained from an intermediate calculation of $\bm \mu_{\bm \beta}$.
\[
\bm \Sigma_{\bm \beta}\bm b=c^{-1}\left(\bm \Sigma_{\bm \beta}\mathbb{X}'\bm \Sigma_{\epsilon}^{-1}\right)_{.,i}.
\]
Here $\left(\bm \Sigma_{\bm \beta}\mathbb{X}'\bm \Sigma_{\epsilon}^{-1}\right)_{.,i}$ is the $i$th column of matrix $\bm \Sigma_{\bm \beta}\mathbb{X}'\bm \Sigma_{\epsilon}^{-1}$ of size $2p\times 2n$.
The mean $\bm \mu_{\bm \beta,-i}$ is
\[
\bm \mu_{\bm \beta,-i}=\bm \Sigma_{\bm \beta,-i}\mathbb{X}_{-i}'(\bm \Sigma_{\epsilon,-i})^{-1}\left[\begin{array}{c}\bm u_{-i} \\ \bm y \end{array},\right],
\]
where $\bm \Sigma_{\bm \beta,-i}$ can be obtained as above, and $(\bm \Sigma_{\epsilon,-i})^{-1}=\left(\bm \Sigma_{\epsilon}^{-1}\right)_{-i,-i}-c^{-1}\left(\bm \Sigma_{\epsilon}^{-1}\right)_{-i,i}\left(\bm \Sigma_{\epsilon}^{-1}\right)_{i,-i}$.

\bibliography{Ref}

@article{wang2007run,
  title={Run-to-run process adjustment using categorical observations},
  author={Wang, Kaibo and Tsung, Fugee},
  journal={Journal of Quality Technology},
  volume={39},
  number={4},
  pages={312--325},
  year={2007},
  publisher={Taylor \& Francis}
}

@article{moustaki2000generalized,
  title={Generalized latent trait models},
  author={Moustaki, Irini and Knott, Martin},
  journal={Psychometrika},
  volume={65},
  number={3},
  pages={391--411},
  year={2000},
  publisher={Springer}
}

@article{liu2014integration,
  title={Integration of data fusion methodology and degradation modeling process to improve prognostics},
  author={Liu, Kaibo and Huang, Shuai},
  journal={IEEE Transactions on Automation Science and Engineering},
  volume={13},
  number={1},
  pages={344--354},
  year={2014},
  publisher={IEEE}
}

@book{shi2006stream,
  title={Stream of variation modeling and analysis for multistage manufacturing processes},
  author={Shi, Jianjun},
  year={2006},
  publisher={CRC press},
  address={Boca Raton}
}

@article{catalano1992bivariate,
  title={Bivariate latent variable models for clustered discrete and continuous outcomes},
  author={Catalano, Paul J and Ryan, Louise M},
  journal={Journal of the American Statistical Association},
  volume={87},
  number={419},
  pages={651--658},
  year={1992},
  publisher={Taylor \& Francis Group}
}

@article{olkin1961multivariate,
  title={Multivariate correlation models with mixed discrete and continuous variables},
  author={Olkin, Ingram and Tate, Robert Fleming and others},
  journal={The Annals of Mathematical Statistics},
  volume={32},
  number={2},
  pages={448--465},
  year={1961},
  publisher={Institute of Mathematical Statistics}
}

@article{cox1992response,
  title={Response models for mixed binary and quantitative variables},
  author={Cox, David R and Wermuth, Nanny},
  journal={Biometrika},
  volume={79},
  number={3},
  pages={441--461},
  year={1992},
  publisher={Oxford University Press}
}

@article{dunson2000bayesian,
  title={Bayesian latent variable models for clustered mixed outcomes},
  author={Dunson, David B},
  journal={Journal of the Royal Statistical Society: Series B (Statistical Methodology)},
  volume={62},
  number={2},
  pages={355--366},
  year={2000},
  publisher={Wiley Online Library}
}

@article{fitzmaurice1995regression,
  title={Regression models for a bivariate discrete and continuous outcome with clustering},
  author={Fitzmaurice, Garrett M and Laird, Nan M},
  journal={Journal of the American statistical Association},
  volume={90},
  number={431},
  pages={845--852},
  year={1995},
  publisher={Taylor \& Francis}
}

@article{zhou2006bayesian,
  title={Bayesian decision procedures for binary and continuous bivariate dose-escalation studies},
  author={Zhou, Yinghui and Whitehead, John and Bonvini, Elisa and Stevens, John W},
  journal={Pharmaceutical Statistics: The Journal of Applied Statistics in the Pharmaceutical Industry},
  volume={5},
  number={2},
  pages={125--133},
  year={2006},
  publisher={Wiley Online Library}
}

@article{cheng2015time,
  title={Time series forecasting for nonlinear and non-stationary processes: a review and comparative study},
  author={Cheng, Changqing and Sa-Ngasoongsong, Akkarapol and Beyca, Omer and Le, Trung and Yang, Hui and Kong, Zhenyu and Bukkapatnam, Satish TS},
  journal={Iie Transactions},
  volume={47},
  number={10},
  pages={1053--1071},
  year={2015},
  publisher={Taylor \& Francis}
}

@article{gueorguieva2001correlated,
  title={A correlated probit model for joint modeling of clustered binary and continuous responses},
  author={Gueorguieva, Ralitza V and Agresti, Alan},
  journal={Journal of the American Statistical Association},
  volume={96},
  number={455},
  pages={1102--1112},
  year={2001},
  publisher={Taylor \& Francis}
}

@article{mcculloch2008joint,
  title={Joint modelling of mixed outcome types using latent variables},
  author={McCulloch, Charles},
  journal={Statistical Methods in Medical Research},
  volume={17},
  number={1},
  pages={53--73},
  year={2008},
  publisher={Sage Publications Sage UK: London, England}
}

@article{hwang2014semiparametric,
  title={Semiparametric Bayesian joint modeling of a binary and continuous outcome with applications in toxicological risk assessment},
  author={Hwang, Beom Seuk and Pennell, Michael L},
  journal={Statistics in medicine},
  volume={33},
  number={7},
  pages={1162--1175},
  year={2014},
  publisher={Wiley Online Library}
}

@article{dunson2003dynamic,
  title={Dynamic latent trait models for multidimensional longitudinal data},
  author={Dunson, David B},
  journal={Journal of the American Statistical Association},
  volume={98},
  number={463},
  pages={555--563},
  year={2003},
  publisher={Taylor \& Francis}
}

@article{yeung2015bayesian,
  title={Bayesian adaptive dose-escalation procedures for binary and continuous responses utilizing a gain function},
  author={Yeung, Wai Yin and Whitehead, John and Reigner, Bruno and Beyer, Ulrich and Diack, Cheikh and Jaki, Thomas},
  journal={Pharmaceutical statistics},
  volume={14},
  number={6},
  pages={479--487},
  year={2015},
  publisher={Wiley Online Library}
}

@article{sun2017functional,
  title={Functional quantitative and qualitative models for quality modeling in a fused deposition modeling process},
  author={Sun, Hongyue and Rao, Prahalad K and Kong, Zhenyu James and Deng, Xinwei and Jin, Ran},
  journal={IEEE Transactions on Automation Science and Engineering},
  volume={15},
  number={1},
  pages={393--403},
  year={2017},
  publisher={IEEE}
}

@article{kang2018bayesian,
  title={A Bayesian hierarchical model for quantitative and qualitative responses},
  author={Kang, Lulu and Kang, Xiaoning and Deng, Xinwei and Jin, Ran},
  journal={Journal of Quality Technology},
  volume={50},
  number={3},
  pages={290--308},
  year={2018},
  publisher={Taylor \& Francis}
}

@article{holmes2006bayesian,
  title={Bayesian auxiliary variable models for binary and multinomial regression},
  author={Holmes, Chris C and Held, Leonhard and others},
  journal={Bayesian analysis},
  volume={1},
  number={1},
  pages={145--168},
  year={2006},
  publisher={International Society for Bayesian Analysis}
}

@article{deng2015QQ,
  title={QQ Models: Joint Modeling for Quantitative and Qualitative Quality Responses in Manufacturing Systems},
  author={Deng, Xinwei and Jin, Ran},
  journal={Technometrics},
  number={3},
  volume={57},
  pages={320--331},
  year={2015},
  publisher={Taylor \& Francis}
}

@article{joseph2006bayesian,
  title={A Bayesian approach to the design and analysis of fractionated experiments},
  author={Joseph, V Roshan},
  journal={Technometrics},
  volume={48},
  number={2},
  pages={219--229},
  year={2006},
  publisher={Taylor \& Francis}
}

@article{kang2009bayesian,
  title={Bayesian optimal single arrays for robust parameter design},
  author={Kang, Lulu and Joseph, V Roshan},
  journal={Technometrics},
  volume={51},
  number={3},
  pages={250--261},
  year={2009}
}

@article{ai2009bayesian,
  title={Bayesian optimal blocking of factorial designs},
  author={Ai, Mingyao and Kang, Lulu and Joseph, V Roshan},
  journal={Journal of Statistical Planning and Inference},
  volume={139},
  number={9},
  pages={3319--3328},
  year={2009},
  publisher={Elsevier}
}

@book{wu2011experiments,
  title={Experiments: planning, analysis, and optimization},
  author={Wu, CF Jeff and Hamada, Michael S},
  volume={552},
  year={2011},
  publisher={John Wiley \& Sons},
  address={New Jersey}
}

@article{LBW1,
  title={Long-term developmental outcomes of low birth weight infants},
  author={Hack, Maureen and Klein, Nancy K and Taylor, H Gerry},
  journal={The future of children},
  pages={176--196},
  year={1995},
  publisher={JSTOR}
}

@article{LBWtrends,
  title={Trends in mortality and morbidity for very low birth weight infants, 1991--1999},
  author={Horbar, Jeffrey D and Badger, Gary J and Carpenter, Joseph H and Fanaroff, Avroy A and Kilpatrick, Sarah and LaCorte, Meena and Phibbs, Roderic and Soll, Roger F and others},
  journal={Pediatrics},
  volume={110},
  number={1},
  pages={143--151},
  year={2002},
  publisher={Am Acad Pediatrics}
}

@article{PTBoverview,
  title={An overview of mortality and sequelae of preterm birth from infancy to adulthood},
  author={Saigal, Saroj and Doyle, Lex W},
  journal={The Lancet},
  volume={371},
  number={9608},
  pages={261--269},
  year={2008},
  publisher={Elsevier}
}

@article{PTB1,
  title={Epidemiology and causes of preterm birth},
  author={Goldenberg, Robert L and Culhane, Jennifer F and Iams, Jay D and Romero, Roberto},
  journal={The lancet},
  volume={371},
  number={9606},
  pages={75--84},
  year={2008},
  publisher={Elsevier}
}

@article{PTBcost,
  title={Cost of hospitalization for preterm and low birth weight infants in the United States},
  author={Russell, Rebecca B and Green, Nancy S and Steiner, Claudia A and Meikle, Susan and Howse, Jennifer L and Poschman, Karalee and Dias, Todd and Potetz, Lisa and Davidoff, Michael J and Damus, Karla and others},
  journal={Pediatrics},
  volume={120},
  number={1},
  pages={e1--e9},
  year={2007},
  publisher={Am Acad Pediatrics}
}

@article{shah2011intention,
  title={Intention to become pregnant and low birth weight and preterm birth: a systematic review},
  author={Shah, Prakesh S and Balkhair, Taiba and Ohlsson, Arne and Beyene, Joseph and Scott, Fran and Frick, Corine},
  journal={Maternal and child health journal},
  volume={15},
  number={2},
  pages={205--216},
  year={2011},
  publisher={Springer}
}

@article{Darrow2009,
  title={Seasonality of birth and implications for temporal studies of preterm birth},
  author={Darrow, Lyndsey A and Strickland, Matthew J and Klein, Mitchel and Waller, Lance A and Flanders, W Dana and Correa, Adolfo and Marcus, Michele and Tolbert, Paige E},
  journal={Epidemiology (Cambridge, Mass.)},
  volume={20},
  number={5},
  pages={699},
  year={2009},
  publisher={NIH Public Access}
}

@article{Palmer2015,
  title={Association between day of delivery and obstetric outcomes: observational study},
  author={Palmer, William L and Bottle, A and Aylin, P},
  journal={Bmj},
  volume={351},
  pages={h5774},
  year={2015},
  publisher={British Medical Journal Publishing Group}
}

\newpage 
\begin{table}
\begin{center}
\caption{The averages and standard errors (in parenthesis) of loss measures when $p = 10$.}
\label{table:p10}
\resizebox{\textwidth}{!}{ 
\begin{tabular}{ccccccccccccccc}
\hline
  \multirow{2}*&&\multicolumn{2}{c}{BLQQ}&&\multicolumn{2}{c}{SM(F)}&&\multicolumn{2}{c}{SM(B)} \\
\cline{3-10}
$\rho$ &      &$s=0.2$  &$s=0.5$ &&$s=0.2$  &$s=0.5$  &&$s=0.2$  &$s=0.5$  \\
\hline
\multirow {5}*{0}
&$RMSE$                         &0.315 (0.021) &0.469 (0.016) &&0.327 (0.016) &0.459 (0.014) &&0.425 (0.014) &0.473 (0.013) \\
&$ME$                           &0.044 (0.005) &0.044 (0.004) &&0.054 (0.004) &0.059 (0.005) &&0.082 (0.005) &0.103 (0.005) \\
&$FSL$                          &0.700 (0.115) &0.500 (0.112) &&7.160 (0.365) &5.640 (0.209) &&0.540 (0.087) &0.420 (0.099) \\
&$L_2(\hat{\bm \beta}_1)$       &9.061 (2.372) &13.19 (1.580) &&18.65 (7.801) &24.53 (8.101) &&8.961 (2.963) &13.34 (2.664) \\
&$L_2(\hat{\bm \beta}_2)$       &0.140 (0.019) &0.247 (0.021) &&0.224 (0.025) &0.333 (0.028) &&0.301 (0.020) &0.349 (0.022) \\
&$\hat{\rho}$                   &0.047 (0.038) &-0.017 (0.030) &&-           &-                &&-           &- \\
\midrule
\multirow {5}*{0.85}
&$RMSE$                         &0.315 (0.018) &0.424 (0.020) &&0.363 (0.022) &0.473 (0.019) &&0.456 (0.015) &0.481 (0.019) \\
&$ME$                           &0.065 (0.004) &0.061 (0.004) &&0.086 (0.004) &0.074 (0.004) &&0.110 (0.005) &0.096 (0.005) \\
&$FSL$                          &0.680 (0.138) &0.740 (0.106) &&5.540 (0.389) &5.680 (0.195) &&0.300 (0.071) &1.160 (0.096) \\
&$L_2(\hat{\bm \beta}_1)$       &3.789 (0.808) &7.425 (1.013) &&37.68 (21.65) &28.64 (9.023) &&10.74 (1.707) &15.72 (3.115) \\
&$L_2(\hat{\bm \beta}_2)$       &0.130 (0.016) &0.267 (0.047) &&0.197 (0.022) &0.378 (0.032) &&0.324 (0.020) &0.395 (0.032) \\
&$\hat{\rho}$                   &0.749 (0.015) &0.785 (0.017) &&-           &-                &&-           &- \\
\midrule
\multirow {5}*{-0.5}
&$RMSE$                         &0.381 (0.026) &0.435 (0.017) &&0.416 (0.022) &0.458 (0.016) &&0.471 (0.019) &0.471 (0.018) \\
&$ME$                           &0.052 (0.004) &0.049 (0.004) &&0.082 (0.004) &0.068 (0.005) &&0.112 (0.004) &0.093 (0.005) \\
&$FSL$                          &0.800 (0.146) &1.040 (0.121) &&6.200 (0.350) &5.420 (0.221) &&0.500 (0.108) &0.960 (0.103) \\
&$L_2(\hat{\bm \beta}_1)$       &3.123 (0.703) &17.98 (5.446) &&26.85 (5.272) &24.35 (6.819) &&8.690 (1.184) &27.34 (6.432) \\
&$L_2(\hat{\bm \beta}_2)$       &0.215 (0.030) &0.230 (0.019) &&0.285 (0.029) &0.317 (0.025) &&0.381 (0.030) &0.336 (0.025) \\
&$\hat{\rho}$                   &-0.512 (0.033) &-0.439 (0.031) &&-           &-                &&-           &- \\
\midrule
\hline
\end{tabular}}
\end{center}
\end{table}

\begin{table}
\begin{center}
\caption{The averages and standard errors (in parenthesis) of loss measures when $p = 30$.}
\label{table:p30}
\resizebox{\textwidth}{!}{ 
\begin{tabular}{ccccccccccccccc}
\hline
  \multirow{2}*&&\multicolumn{2}{c}{BLQQ}&&\multicolumn{2}{c}{SM(F)}&&\multicolumn{2}{c}{SM(B)} \\
\cline{3-10}
$\rho$ &      &$s=0.2$  &$s=0.5$ &&$s=0.2$  &$s=0.5$  &&$s=0.2$  &$s=0.5$  \\
\hline
\multirow {5}*{0}
&$RMSE$                         &0.647 (0.030) &0.804 (0.023) &&0.633 (0.023) &0.842 (0.021) &&0.896 (0.025) &0.936 (0.021) \\
&$ME$                           &0.087 (0.005) &0.161 (0.008) &&0.094 (0.006) &0.148 (0.007) &&0.130 (0.005) &0.144 (0.006) \\
&$FSL$                          &2.800 (0.206) &5.640 (0.298) &&19.60 (0.648) &16.98 (0.483) &&2.720 (0.216) &5.220 (0.332) \\
&$L_2(\hat{\bm \beta}_1)$       &11.59 (0.847) &67.07 (4.089) &&17.03 (3.243) &81.64 (3.192) &&20.79 (0.389) &89.91 (0.911) \\
&$L_2(\hat{\bm \beta}_2)$       &0.605 (0.055) &0.874 (0.052) &&0.600 (0.046) &1.075 (0.059) &&1.376 (0.068) &1.368 (0.060) \\
&$\hat{\rho}$                   &-0.027 (0.040) &0.017 (0.041) &&-           &-                &&-           &- \\
\midrule
\multirow {5}*{0.85}
&$RMSE$                         &0.671 (0.028) &0.761 (0.019) &&0.702 (0.031) &0.865 (0.019) &&0.926 (0.025) &0.906 (0.020) \\
&$ME$                           &0.091 (0.006) &0.141 (0.006) &&0.115 (0.005) &0.181 (0.005) &&0.135 (0.005) &0.180 (0.005) \\
&$FSL$                          &2.100 (0.210) &5.020 (0.302) &&17.86 (0.794) &16.14 (0.472) &&3.320 (0.247) &10.12 (0.309) \\
&$L_2(\hat{\bm \beta}_1)$       &24.35 (1.555) &64.11 (2.646) &&26.51 (1.617) &79.71 (3.457) &&43.79 (0.470) &90.35 (0.620) \\
&$L_2(\hat{\bm \beta}_2)$       &0.583 (0.049) &0.768 (0.050) &&0.769 (0.072) &1.160 (0.057) &&1.507 (0.082) &1.278 (0.060) \\
&$\hat{\rho}$                   &0.801 (0.017) &0.737 (0.018) &&-           &-                &&-           &- \\
\midrule
\multirow {5}*{-0.5}
&$RMSE$                         &0.630 (0.032) &0.831 (0.026) &&0.729 (0.026) &0.851 (0.024) &&0.899 (0.023) &0.886 (0.022) \\
&$ME$                           &0.110 (0.006) &0.148 (0.006) &&0.129 (0.006) &0.154 (0.005) &&0.147 (0.005) &0.158 (0.005) \\
&$FSL$                          &2.240 (0.023) &7.600 (0.204) &&21.70 (0.647) &16.66 (0.569) &&2.500 (0.273) &11.82 (0.235) \\
&$L_2(\hat{\bm \beta}_1)$       &12.64 (1.204) &56.02 (2.578) &&15.37 (1.172) &72.65 (2.174) &&21.03 (0.421) &75.36 (0.742) \\
&$L_2(\hat{\bm \beta}_2)$       &0.564 (0.056) &0.952 (0.064) &&0.823 (0.080) &1.180 (0.067) &&1.353 (0.081) &1.308 (0.062) \\
&$\hat{\rho}$                   &-0.463 (0.028) &-0.446 (0.023) &&-           &-                &&-           &- \\
\midrule
\hline
\end{tabular}}
\end{center}
\end{table}

\begin{table}
\centering
\caption{The variables used in the birth records case study.}
\label{tab:case}
{\small
\begin{tabular}{|p{3cm}| p{6cm}| p{5.8cm}|}
\hline
Variable name & Variable description & Type of variable \\
\hline
z: preterm Birth & Indicator variable for whether the child was born preterm (defined as born before 36 gestational weeks) & Dichotomous dependent variable (1 = preterm, 0 = non preterm)\\
y: Birth Weight & Weight of the infant at birth in grams & Quantitative dependent variable \\
$x_1$: Day of birth & Day of the year (1-366) the infant was born & Quantitative independent variable \\
$x_2$: Day of week & Whether the infant was born on a weekend or a weekday & Dichotomous independent variable (1 = weekend, 0 = weekday) \\
$x_3$: Age of mother & Age of the mother in years & Quantitative independent variable \\
$x_4$: Race & Race reported on the birth record collapsed to whether the infant is identified as African-American  or not & Dichotomous independent variable (1 = African-American, 0 = Not African-American) \\
$x_5$: Ethnicity & Whether the infant is identified as Hispanic or not & Dichotomous independent variable (1 = Hispanic, 0 = Not Hispanic) \\
$x_6$: Mother's Education & Whether the mother completed at least high school or not & Dichotomous independent variable (1 = More than High School, 0 = High School or less) \\
$x_7$: Marriage status & Whether the mother was married at the time of birth or not & Dichotomous independent variable (1 = Married, 0 = Not Married) \\
$x_8$: Sex of child & The sex of the infant & Dichotomous independent variable (1 = Male infant, 0 = Female infant) \\
$x_9$: Parity & Number of pregnancies carried to 24 weeks gestation collapsed to whether this is the first such pregnancy or not & Dichotomous independent variable (1 = First pregnancy, 0 = Not first pregnancy)\\
\hline
\end{tabular}
}
\end{table}

\end{document}